# HW Boötis: an enigmatic cataclysmic variable star

Jeremy Shears, David Boyd, Graham Darlington, Ian Miller, Roger Pickard, Gary Poyner, Richard Sabo, Mike Simonsen & William Stein

**Abstract**

We present the 13-year light curve of HW Boo between 2001 May and 2014 May. We identified 12 outbursts, which typically lasted 2 to 5 days, with an amplitude of 2.7 to 3.6 magnitudes. Time resolved photometry during two outbursts revealed small hump-like structures which increased in size as the outburst progressed, reaching a peak-to-peak amplitude of ~0.8 mag. They occurred on timescales of 15 min to an hour, but did not exhibit a stable period. Similar irregular hump-like variations of 0.1 to 0.8 magnitudes, at intervals of 7 to 30 minutes, were also detected during quiescence. We discuss whether HW Boo might be a dwarf nova of the SU UMa family or an Intermediate Polar, but require further observations to support classification.

**Introduction**

Cataclysmic variables (CVs) are semi-detached binaries in which a white dwarf primary accretes material from a secondary star via Roche lobe overflow. The secondary is usually a late-type main-sequence star. In the absence of a significant white dwarf magnetic field, material from the secondary is processed through an accretion disc before settling on the surface of the white dwarf. In CVs with low to intermediate mass transfer rates, dwarf nova outbursts with amplitudes of 2–8 mag and durations of days to weeks are observed. These outbursts are thought to be caused by a thermal instability in the accretion disc. In cases where the white dwarf has a stronger magnetic field, the inner portion of the disc becomes truncated and the accreting material follows the magnetic field lines and falls onto the poles of the white dwarf. Such systems are called Intermediate Polars (IPs). IPs also exhibit short-term variability mostly associated with turbulence in the accretion system. For more information about CVs, dwarf novae and IPs, the reader is directed to reference (1) and (2).

HW Boo was identified as a dwarf nova by Aungwerojwit et al. (3) in their study of optically selected CV candidates from the Hamburg Quasar Survey (4) (catalogue reference HS 1340+1524). They reported quiescence spectroscopy showing strong single-peaked line profiles of Balmer emission along with weaker lines of He I and He II. The spectroscopic period was 92.66(17) min (0.0643(1) d), which they interpreted as the orbital period, $P_{orb}$. Time resolved photometry using a variety of telescopes showed a mean magnitude in the range ~17.7 – 16.8. In 2003 April they observed a short dwarf nova-like outburst of less than 2 days duration in which the star reached V ≈ 14.7. Spectroscopy showed weak emissions in Hα, whereas Hβ and Hγ were in absorption with narrow emission cores, consistent with an optically thick accretion disc. They recorded a second outburst to V ≈ 14.2 in 2005 April, again with the duration of about 2 -3 days. They reported substantial short-term variability during quiescence, predominantly on time scales of ~15 to 20 min with peak-to-peak amplitudes of ~0.4 mag. On some occasions hump-like structures were found onto which were superimposed short time scale flickering. No stable period was identified in the combined photometry. HW Boo was also catalogued in the Sloan Digital Sky Survey as SDSS J134323.16+150916.8 (5).

In this paper we report our analysis of 13 years of photometry of HW Boo as well as time resolved photometry conducted during two outbursts, one in 2009 and one in 2014.

## Observations

*Long-term photometry*

To produce a long-term light curve of HW Boo, we used data from the AAVSO International Database, which contains more than 3800 observations of the star comprising both visual estimates and CCD measurements, having removed the “fainter than” observations. In addition we used some 345 data points from the Catalina Real-Time Transient Survey (CRTS) (6) and data from Aungwerojwit et al. (3).

*Time resolved photometry*

We obtained 23 h of photometry during the 2009 outburst of HW Boo, 46 h during the 2014 outburst and a further 18 h some 3 weeks after the outburst when the system had reached quiescence. Our instrumentation is shown in Table 1 and the observation log in Table 2. Images were dark-subtracted and flat-fielded prior to being measured using differential aperture photometry relative to the AAVSO V-band sequence (7).

## Results

*Long-term light curve*

We present the 13-year light curve of HW Boo, from 2001 May to 2014 May, in Figure 1. We identified 12 outbursts, taking an outburst as a brightening to magnitude 16.0 or less, the details of which are given in Table 3. The outbursts lasted 2 to 5 days and the maximum observed brightness during outburst was between magnitude 13.4 and 15.8, with an average of 14.8. The fainter maxima might simply reflect poor data sampling, thereby missing the actual maximum. The time between recorded outbursts varies between 34 and 735 days, with an average of 366 days. However, given the relatively low cadence of the observations, especially during the first few years until CRTS monitoring began, as well as the seasonal gaps on observations, is it is probable that we have missed some outbursts. The median interval between observations (not including seasonal gaps) was 1.0 days and the mean was 3.2 days, which is comparable to the low limit of outburst duration.

The mean quiescent magnitude was 16.8. There was considerable variation in quiescent brightness, with a standard deviation of 0.5 and a minimum of 18.5. Whilst some of this apparent scatter might be explained by errors inherent in the measurements, the majority can be attributed to real variations in brightness. Later in the paper we shall present our measurements of short terms variations, which are consistent with the ~0.4 magnitude variations in quiescence reported by Aungwerojwit et al. (3).

Careful examination of Figure 1 also shows an apparent long-term evolution in quiescence brightness. This is even more evident when one considers the CRTS data separately as shown in the inset to Figure 1. We highlight the CRTS data since they were obtained with a single telescope resulting in a self-consistent data set, with a mean error of 0.08 mag. Fitting a third order polynomial to the CRTS quiescent data, i.e. having eliminated the outbursts, (shown as the line in Figure 1 inset) to guide the eye shows that the star gradually increased in brightness from mean a quiescent magnitude at the beginning of the CRTS observations of ~17.4 in 2005 March to ~16.7 in early 2012 after which it begins to fade.

*Comparisons of the 2003, 2009, 2012 May and 2014 outbursts*

Expanded light curves of the four best observed outbursts are shown in Figure 2. The maxima were magnitude 14.7, 13.5, 14.3 and 14.6 respectively, confirming that the amplitude of outbursts is not constant. To investigate the amplitude further, we established a local mean quiescent magnitude in the 25 days before the outburst, thereby eliminating the effect of long term evolution in quiescent brightness, resulting in amplitudes of 3.6, 2.7 and 3.2 for the 2009, 2012 and 2014 outbursts. There were insufficient data to perform this analysis for 2003, but Aungwerojwit et al. (3) report a quiescent magnitude of 17.6 shortly after the outburst, implying an outburst amplitude of 2.9 magnitudes.

Superficially there is some similarity in the outburst profile among the four outbursts. Although the rises to outburst were not caught, in general it was fast, with slower declines, as is typical of dwarf novae. Apparently, little time was spent at maximum, with a total outburst duration of 3 to 5 days.

*Time resolved photometry*

Figures 3 and 4 show some of the longer photometry runs obtained during the 2009 and 2014 outbursts plotted to the same scale. During the initial stages of the fade in both outbursts small hump-like structures were observed which increased in size during the rest of the outburst, reaching a peak-to-peak amplitude of ~0.8 mag. Superimposed on these was small-scale flickering. We analysed the combined datasets for each outburst, as well as the daily time resolved photometry runs, using the Lomb-Scargle and ANOVA algorithms in the Peranso V2.50 software (8), but could not find a significant stable period in the power spectra, including in the region of the proposed $P_{orb}$ (data not shown). Visual inspection of the light curves shows that the humps occur in an irregular manner with an interval between them of 15 mins to about an hour.

Photometry conducted some 3 weeks after the outburst, when the system had reached quiescence, also showed large (up to 0.8 mag) variations (Figure 5 – note the light curves are plotted on an expanded JD scale compared to those in Figure 4). Again period analysis showed these had no significant period associated with them. The night-to-night variations were also remarkable. On JD 2456791 (Figure 5, middle plot), fifteen humps were observed during 4.2 h of photometry; these had amplitudes between 0.1 and 0.5 mag and occurred at intervals of ~8 to 30 minutes. By contrast, the following night, JD 2456791 (Figure 5, bottom plot), a rapid fade of 0.8 mag was recorded at the beginning of the photometric run. The rest of the run was characterised by 0.1 to 0.3 mag humps at irregular intervals of ~ 7 to 20 minutes.

Our photometric results showing irregular variations are similar to those reported by Aungwerojwit et al. (3) at the time of the 2003 outburst.

## Discussion

Aungwerojwit et al. (3) indentified HW Boo as a likely dwarf nova, given their detection of two outbursts of this CV. Their $P_{orb}$ of 92.66(17) min places it well below the so-called period gap of 2 to 3 h in the period distribution of CVs. The SU UMa family of dwarf novae dominates such territory. SU UMa systems occasionally exhibit superoutbursts that last several times longer than normal outbursts and may be up to a magnitude brighter; whereas normal outbursts last a few days, superoutbursts typically last one to two weeks. During a superoutburst the light curve of a SU UMa system is characterised by superhumps. These are modulations in the light curve which are a few percent longer than the orbital period. They are thought to arise from the interaction of the secondary star orbit with a slowly precessing eccentric accretion disc. The eccentricity of the disc arises because a 3:1 resonance occurs between the secondary star orbit and the motion of matter in the outer accretion disc.

The twelve outbursts of HW Boo we report here are consistent with normal outbursts of a dwarf nova. They were short in duration and none had the prolonged plateau that would be typical of a superoutburst. Moreover superhumps, the defining visible manifestation of a SU UMa system, were absent. However, the fact that we have not observed a superoutburst does not mean that HW Boo is not a member of the SU UMa family. It may be that we have missed one or more superoutbursts due to the observational gaps in our 13-year coverage. Or it could be that the superoutburst frequency is low. There are possible parallels here with the SU UMa star, BZ UMa, whose $P_{orb}$ of 97.8 min (9) is similar to HW Boo's. BZ UMa was monitored for 40 years, during which 28 normal outbursts were observed, before an elusive superoutburst was finally confirmed (10).

The long-term light curve of HW Boo is also reminiscent of IPs, a few of which show outbursts of similar amplitude, duration and frequency (indeed before its SU UMa status was confirmed it was suggest by some that BZ UMa might be an IP (10)). For example, TV Col has shown outbursts of up to two magnitudes over a period of a few hours (11). DO Dra showed 5 outbursts between 1996 and 2004, each lasting about 5 days and with an amplitude of 4 to 5 mag (12). EX Hyi undergoes ~3.5 mag outbursts lasting 2- 3 d every ~1.5 years (13). Other outbursting IPs include XY Ari (14), GK Per (15) and V1223 Sgr (16). Moreover, IPs are X-ray sources and HW Boo was detected as an X-ray source in the ROSAT All Sky Survey (1RXS J134323.1+150916) (3). However, dwarf novae also emit X-rays so this is not diagnostic. Nevertheless the X-ray emission has a hardness ratio HR1 = 0.18(23), which is much softer than most non-magnetic CVs were HR ≈ 1 (3).

HW Boo exhibits some other unusual characteristics that are not uniquely associated with a dwarf nova or IP classification. The large amplitude and rapid hump-like variations during the decline from outburst and in quiescence are reminiscent of the flaring or Quasi-Periodic Oscillations (QPOs) seen in several IPs, such as EX Hyi during its 1987 outburst (17). Conversely, they are also seen in some dwarf novae, including BZ UMa which was found to exhibit a 0.3 mag QPO-like phenomenon with a period around 20 min during its 1999 March outburst (18).

The long-term evolution of the average quiescent brightness is another unusual feature of HW Boo's light curve. Whilst most dwarf novae appear to have a relatively constant quiescent magnitude over several years, some, such as FO And (19) (SU UMa family) and HT Cam (an IP) (20), appear to show a similar evolution to HW Boo. Whether this reflects changes in accretion rate, changes in the secondary star irradiation, or another factor is currently unknown.

Only further observations will resolve the question of whether HW Boo is an SU UMa-type dwarf nova, an IP, or some other form of CV. Long term monitoring will be important to determine its true outburst frequency and to identify a potential superoutburst. This should trigger time resolved photometry to look for superhumps, which would be diagnostic of its SU UMa identity. Amateur astronomers, suitably equipped, are ideally placed to contribute to these activities. Similarly further work, including polarimetry to look for circular polarisation and X-ray observations to look for X-ray pulses, may shed light on the possible IP explanation. However, as more CVs are studied, it is becoming apparent that the neat classification into distinct classes might not be quite so straightforward. A case in point is QZ Vir (T Leo) which Vrielmann et al. (21) suggest could be a superhumping IP!

## Conclusions

The 13-year light curve of HW Boo between 2001 May and 2014 May reveals 12 outbursts, which typically lasted 2 to 5 days, with an amplitude of 2.7 to 3.6 magnitudes above quiescence. Time resolved photometry during two outbursts revealed small hump-like structures in the light curve which increased in size as the outburst progressed, reaching a peak-to-peak amplitude of ~0.8 mag. Similar irregular hump-like variations of 0.1 to 0.8 magnitudes, at intervals of 7 to 30 minutes, were also detected during quiescence. Whilst HW Boo has some properties consistent with dwarf novae of the SU UMa family, we suggest that more likely it is an Intermediate Polar. We encourage further observations to support its classification.

## Acknowledgments

The authors gratefully acknowledge the use of observations from the AAVSO International Database contributed by observers worldwide. We are indebted to Professor Boris Gänsicke, University of Warwick, UK, for allowing us to use the original photometry reported in Aungwerojwit et al. (3). GP thanks the AAVSO for the use of the Sonoita Research Observatory (SRO), part of the AAVSOnet virtual observatory, and the Department of Cybernetics at the University of Bradford, UK, for the use of the Bradford Robotic Telescope (BRT), located at the Teide Observatory on Tenerife in the Canary Islands. This research made use of data from the Catalina Real-Time Transient Survey. We also used SIMBAD, operated through the Centre de Données Astronomiques (Strasbourg, France) and the NASA/Smithsonian Astrophysics Data System. We thank our referees, Dr Chris Lloyd and Dr. Elmé Breedt, for their helpful comments.

## Addresses

Shears: “Pemberton”, School Lane, Bunbury, Tarporley, Cheshire, CW6 9NR, UK [bunburyobservatory@hotmail.com]

Boyd: 5 Silver Lane, West Challow, Wantage, Oxon, OX12 9TX, UK [davidboyd@orion.me.uk]

Darlington: Thakeham Observatory, West Sussex, UK [graham@midnight-blue.org.uk]

Miller: Furzehill House, Ilston, Swansea, SA2 7LE, UK [furzehillobservatory@hotmail.com]

Pickard: 3 The Birches, Shobdon, Leominster, Herefordshire. HR6 9NG, UK [roger.pickard@sky.com]

Poyner: 67 Ellerton Road, Kingstanding, Birmingham, B44 0QA [garypoyner@blueyonder.co.uk]

Sabo: 2336 Trailcrest Drive, Bozeman, MT 59718, USA [rsabo333@gmail.com]

Simonsen: AAVSO, 49 Bay State Rd. Cambridge MA 02139, USA [mikesimonsen@aavso.org]

Stein: 6025 Calle Paraiso, Las Cruces, NM 88012, USA [starman@tbelc.org]

| Observer | Telescope | CCD |
|---|---|---|
| Boyd | 0.35m SCT | Starlight Xpress SXV-H9 |
| Darlington | 0.35m SCT | SBIG ST9-XE |
| Miller | 0.35m SCT | Starlight Xpress SXVR-H16 |
| Pickard | 0.35m SCT | Starlight Xpress SXVF-H9 |
| Sabo | 0.43m reflector | SBIG STL-1001 |
| Shears | 0.28m SCT | Starlight Xpress SXVF-H9 |
| Simonsen | 0.3 m SCT | SBIG ST9 |
| Stein | 0.35m SCT | SBIG ST10XME |

**Table 1: Instrumentation**

| Date (UT) | Start time | End time | Duration (h) | Observer |
|---|---|---|---|---|
| **2009** | | | | |
| April 16 | 2454937.729 | 2454937.798 | 1.7 | Simonsen |
| April 18 | 2454940.357 | 2454940.437 | 1.9 | Shears |
| April 19 | 2454940.664 | 2454940.862 | 4.8 | Stein |
| April 19 | 2454941.363 | 2454941.441 | 1.9 | Pickard |
| April 20 | 2454941.650 | 2454941.858 | 5.0 | Stein |
| April 21 | 2454942.850 | 2454941.862 | 0.3 | Stein |
| April 21 | 2454942.830 | 2454941.952 | 2.9 | Sabo |
| April 22 | 2454943.681 | 2454943.853 | 4.1 | Stein |
| **2014** | | | | |
| April 18 | 2456766.344 | 2456766.466 | 2.9 | Boyd |
| April 18 | 2456766.356 | 2456766.508 | 3.6 | Shears |
| April 18 | 2456766.368 | 2456766.647 | 6.8 | Miller |
| April 18 | 2456766.477 | 2456766.631 | 3.7 | Pickard |
| April 19 | 2456766.720 | 2456766.957 | 5.7 | Sabo |
| April 19 | 2456767.364 | 2456767.645 | 6.7 | Miller |
| April 19 | 2456767.457 | 2456767.625 | 4.0 | Pickard |
| April 21 | 2456768.651 | 2456768.938 | 6.9 | Sabo |
| April 21 | 2456768.698 | 2456768.802 | 2.5 | Stein |
| April 22 | 2456769.683 | 2456769.710 | 0.6 | Stein |
| April 22 | 2456770.430 | 2456770.585 | 2.5 | Darlington |
| May 12 | 2456790.381 | 2456790.525 | 3.5 | Darlington |
| May 13 | 2456791.419 | 2456791.595 | 4.2 | Pickard |
| May 13 | 2456791.449 | 2456791.573 | 3.8 | Miller |
| May 14 | 2456792.413 | 2456792.569 | 3.7 | Pickard |
| May 14 | 2456792.448 | 2456792.570 | 2.9 | Miller |

**Table 2: Log of time resolved photometry**

| Date of outburst | JD | Maximum brightness (mag) | Duration (d) | Time since previous outburst (d) |
|---|---|---|---|---|
| 2003 April 11 | 2452740 | 14.7 | 3 to 4 | |
| 2005 Apr 15 | 2453476 | 14.2 | 2 to 3 | 735 |
| 2006 January 28 | 2453764 | 15.8 | <5 | 288 |
| 2007 March 21 | 2454181 | 15.5 | <5 | 417 |
| 2007 April 24 | 2454215 | 15.0 | 2 to 3 | 34 |
| 2009 April 17 | 2454939 | 13.4 | 4 to 5 | 725 |
| 2010 July 10 | 2455388 | 14.7 | 4 to 5 | 448 |
| 2011 March 26 | 2455647 | 14.8 | >1 | 259 |
| 2012 January 26 | 2455953 | 14.5 | 3 to 4 | 306 |
| 2012 May 15 | 2456063 | 14.7 | 2 to 3 | 111 |
| 2013 April 10 | 2456393 | 15.2 | 2 to 3 | 329 |
| 2014 April 18 | 2456766 | 14.6 | 3 to 4 | 373 |

**Table 3: Recorded outbursts of HW Boo**

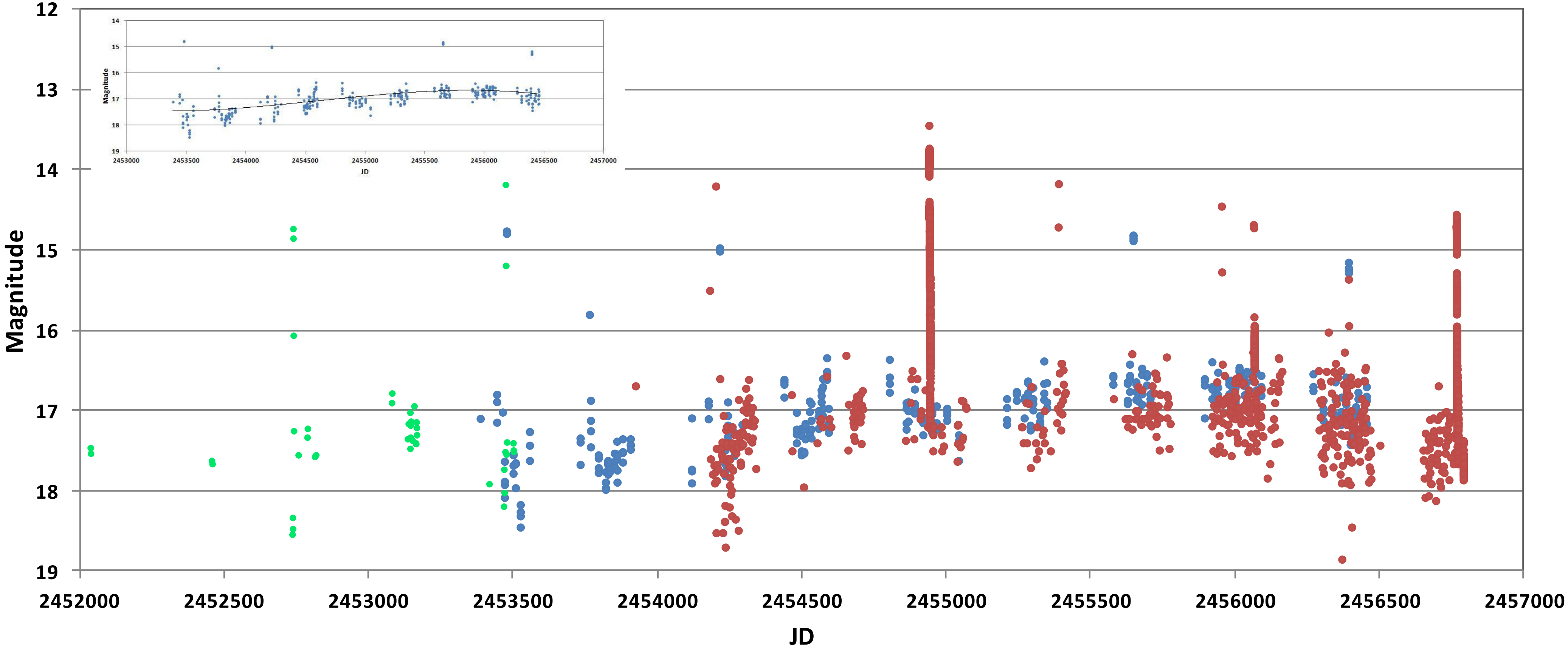

**Figure 1: The thirteen year light curve of HW Boo from 2001 May to 2014 May. Main plot data points: red is AAVSO International Database, blue is CRTS, green is Aungwerojwit et al.** (3)**. Inset is a plot showing CRTS data only with a third order polynomial fit to the quiescent data**

**Figure 2: Outburst light curves**

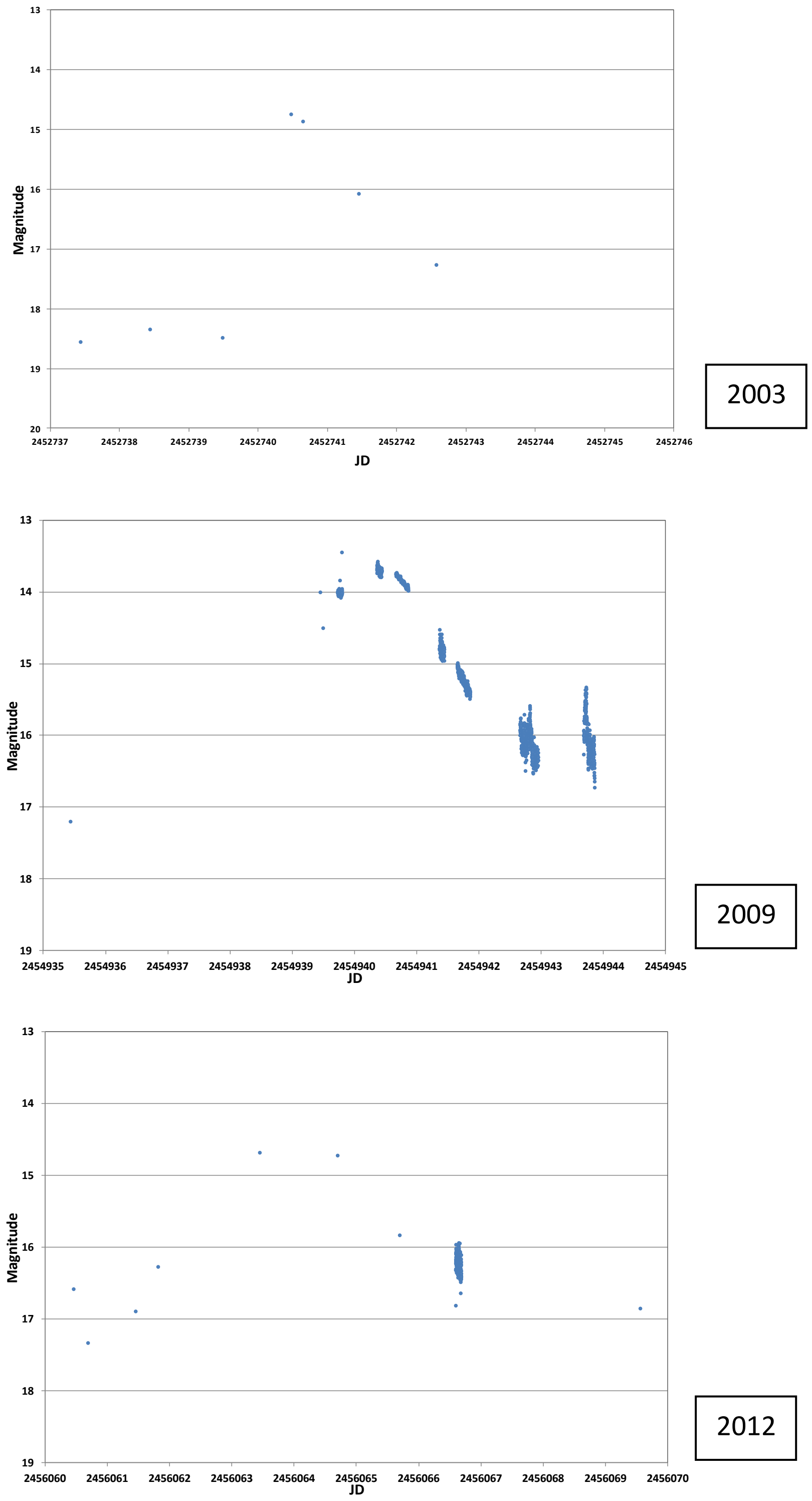

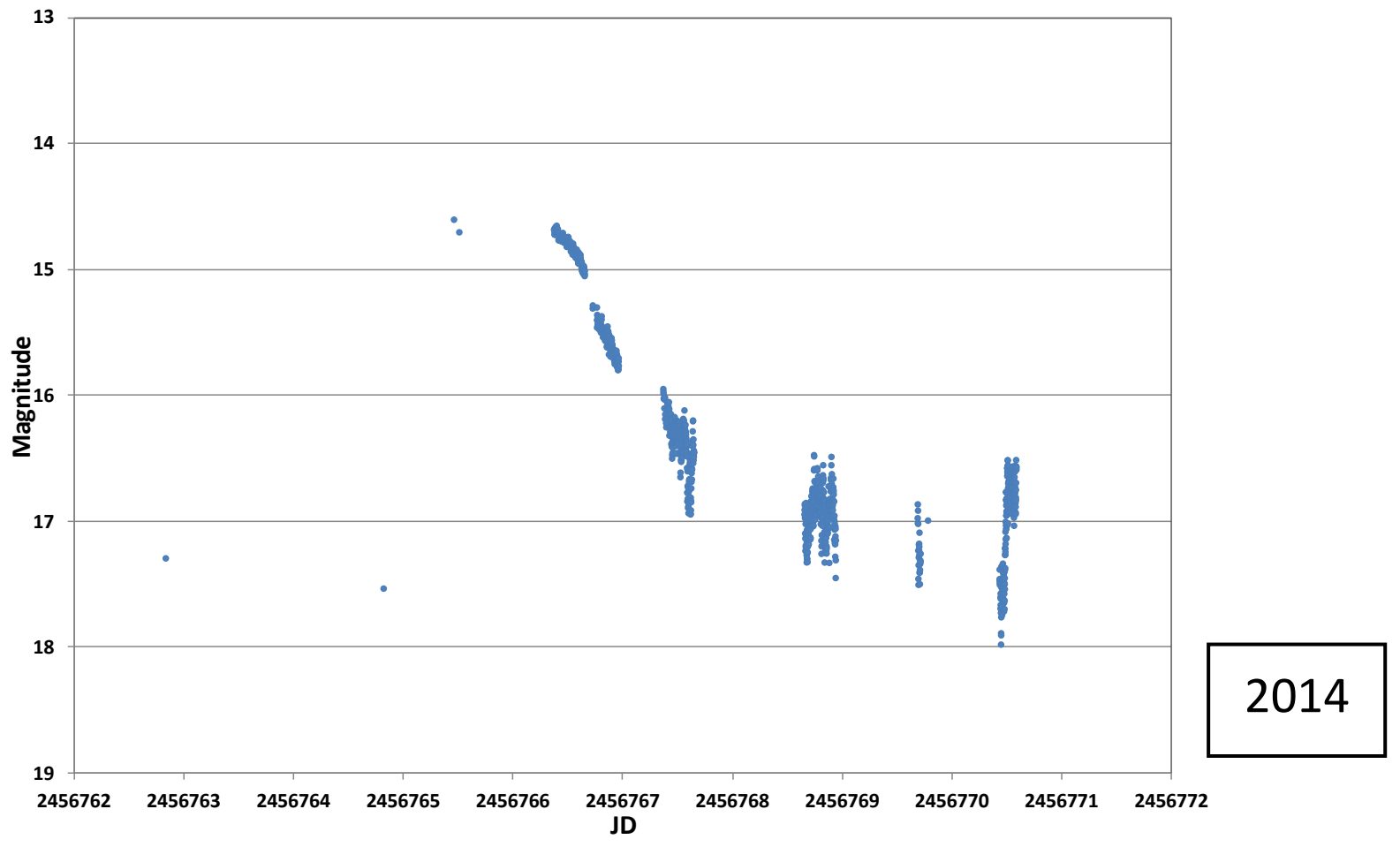

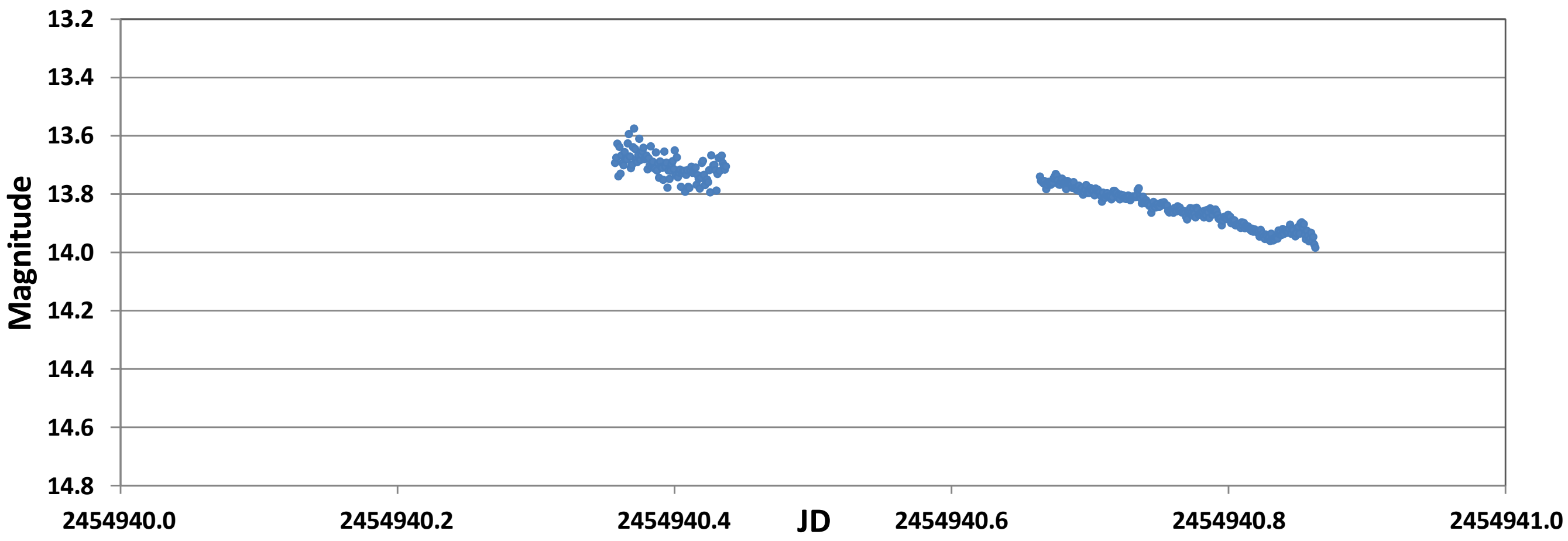

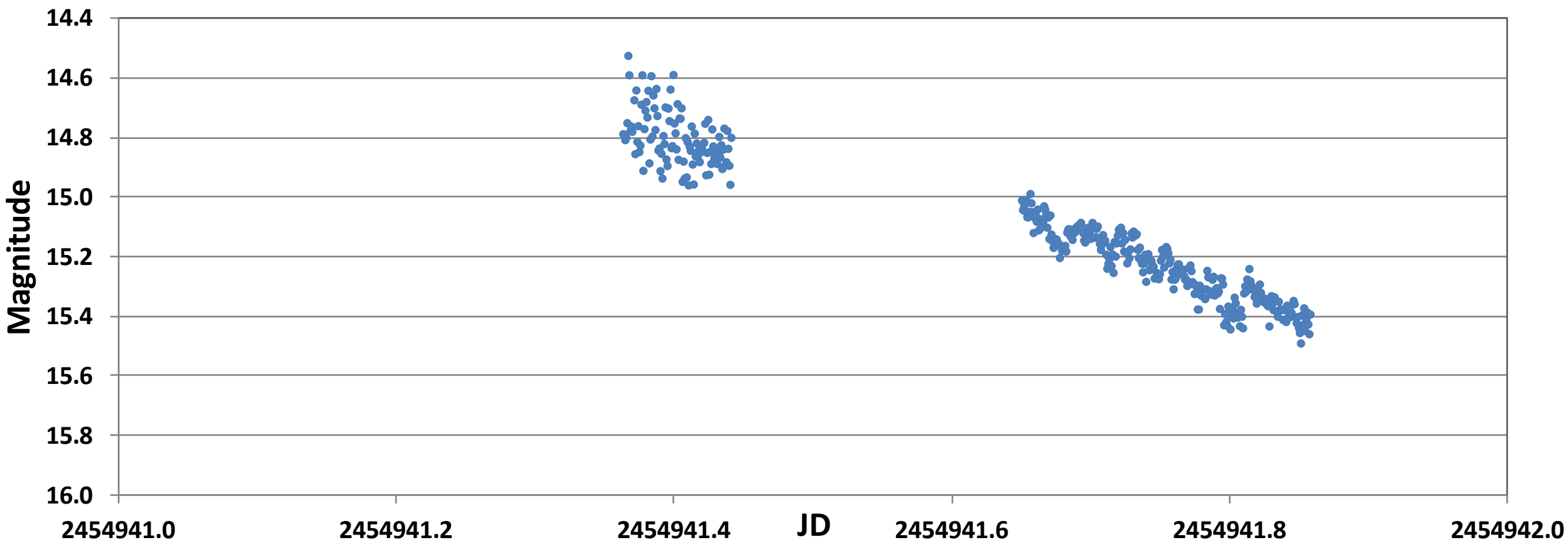

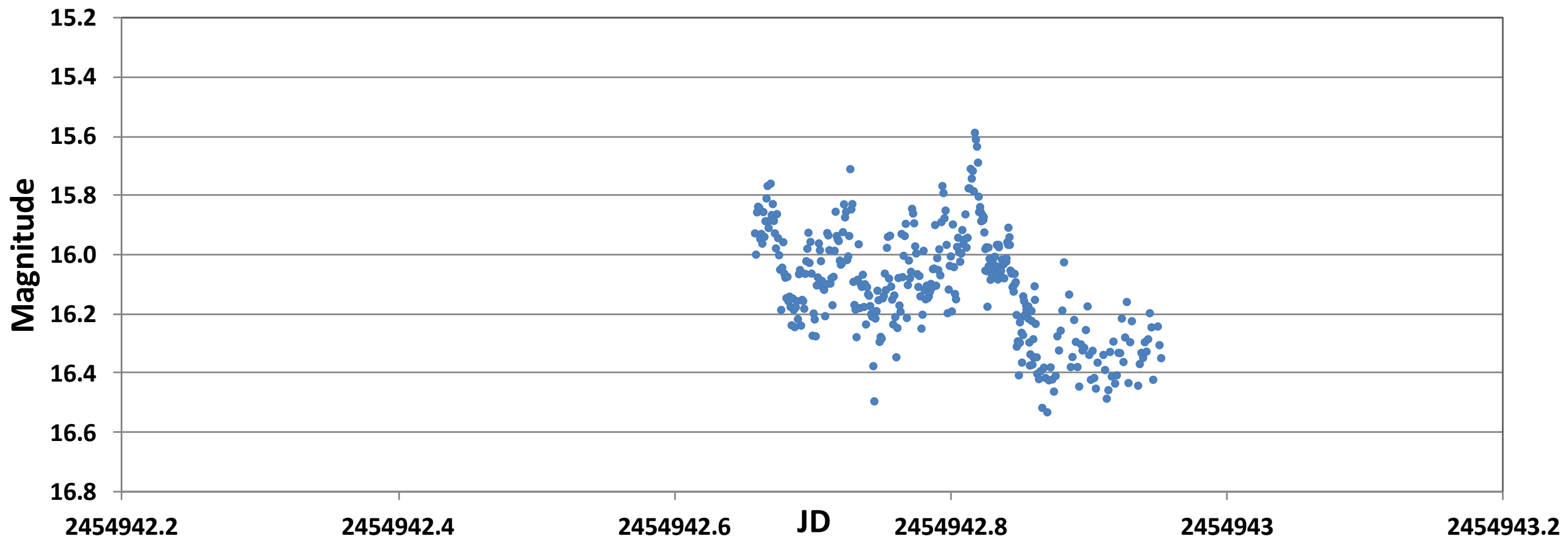

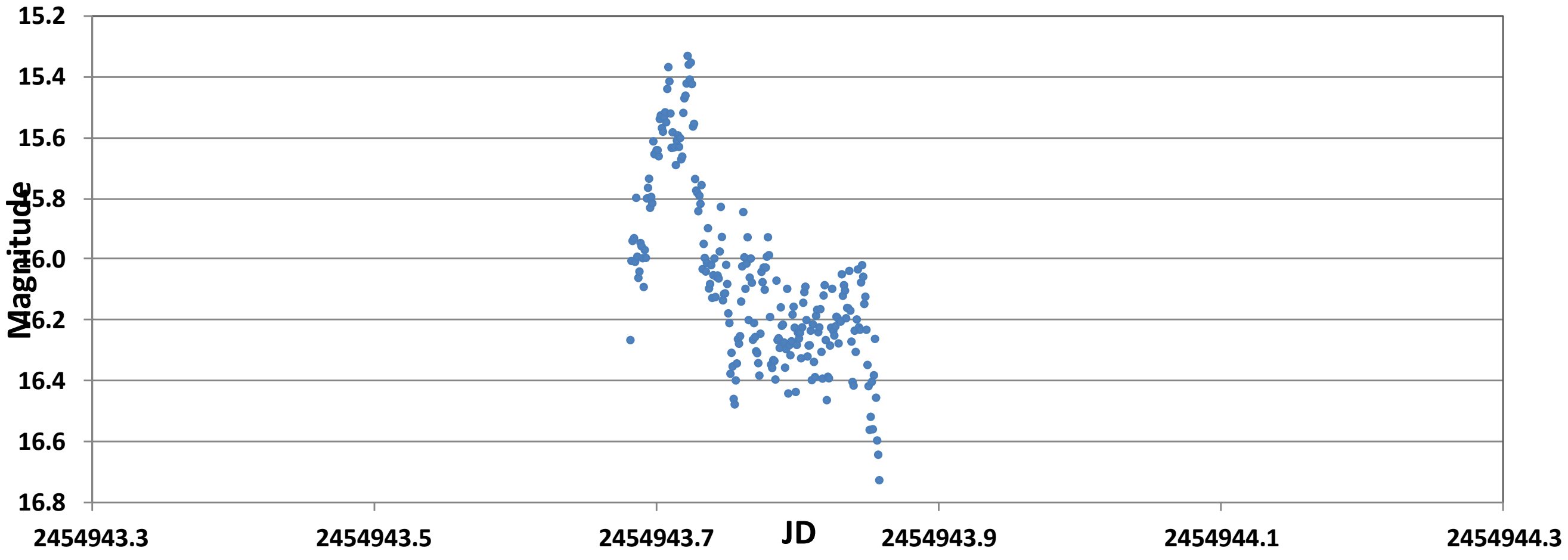

**Figure 3: Time resolved photometry during the 2009 outburst**

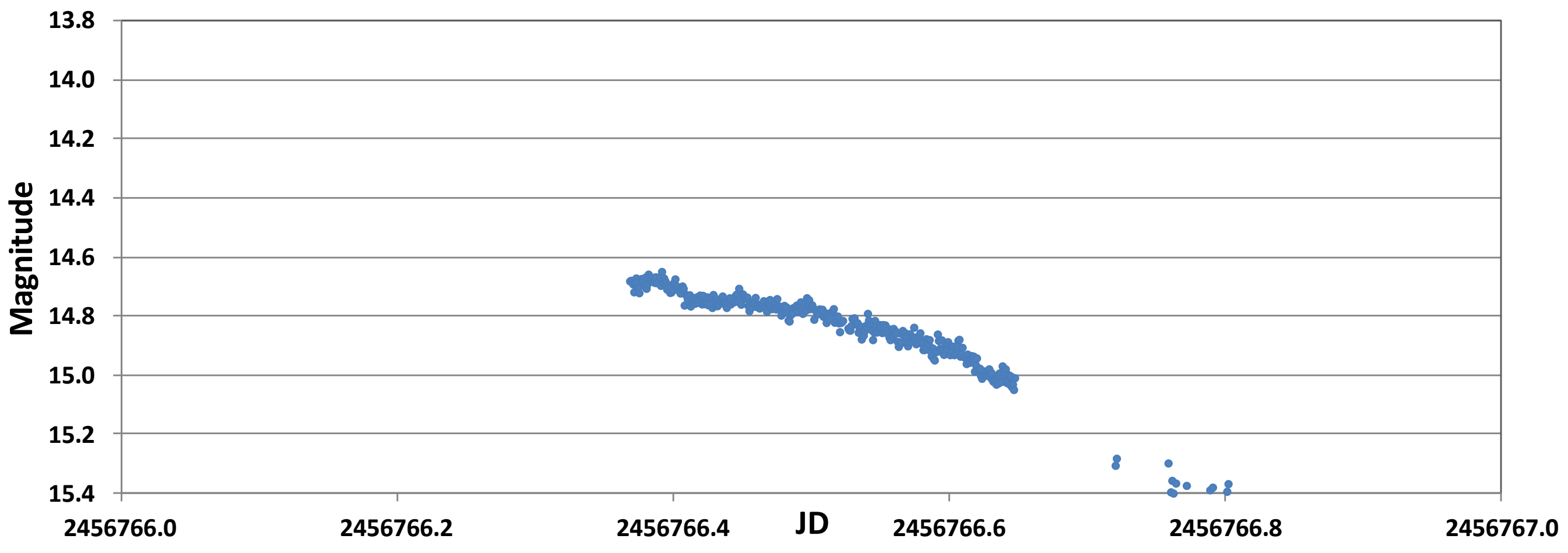

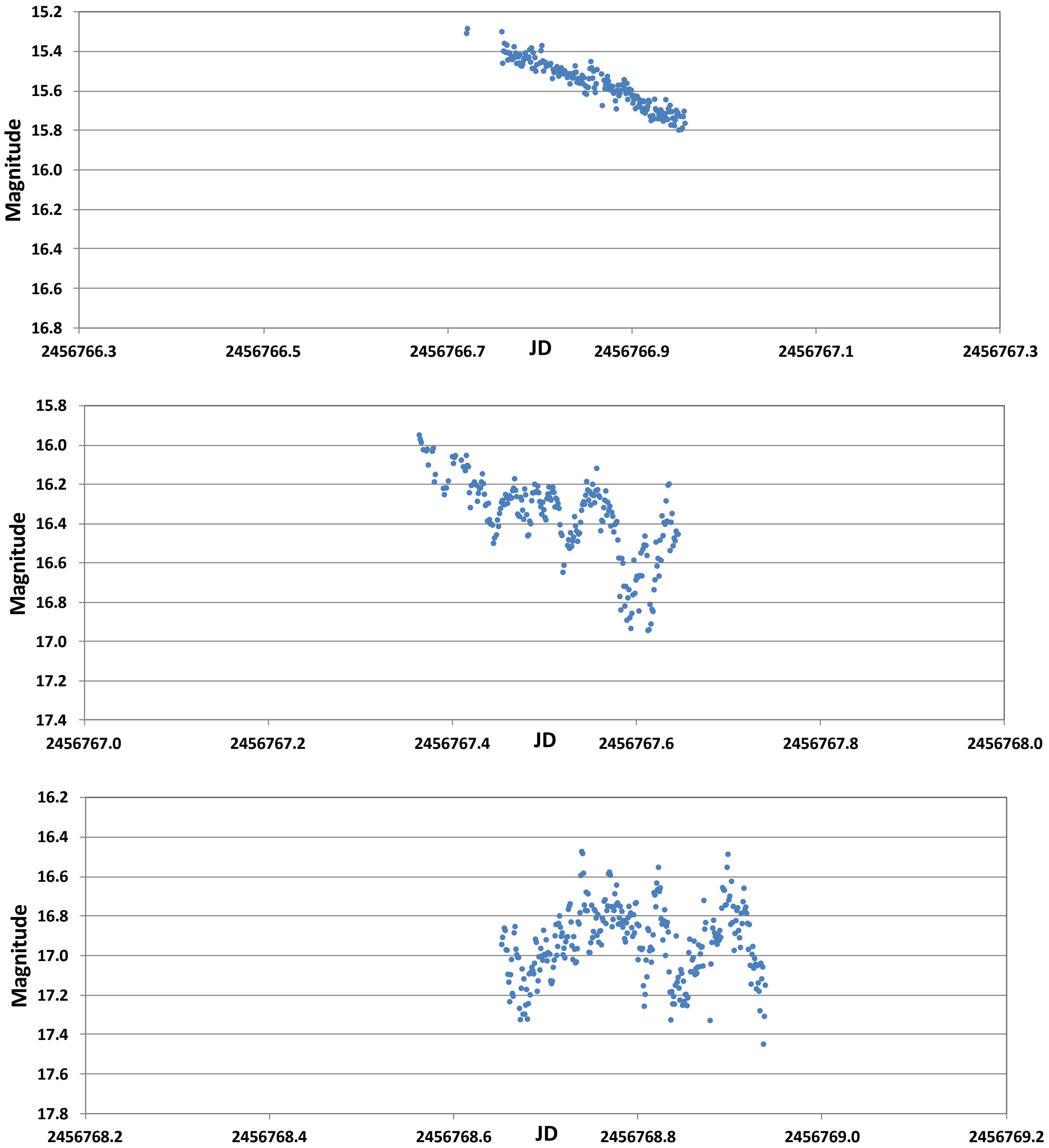

**Figure 4: Time resolved photometry during the 2014 outburst**

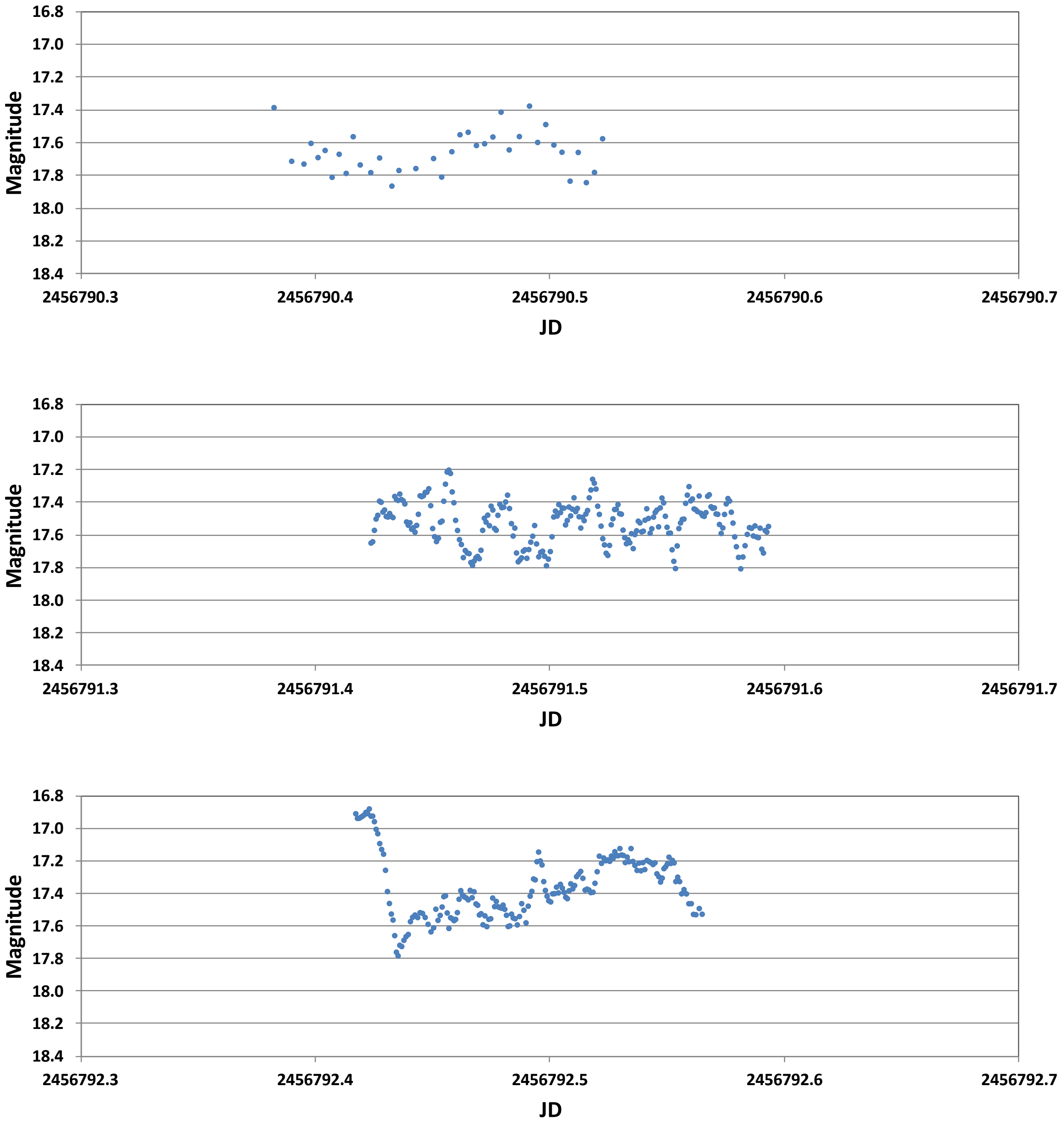

**Figure 5: Time resolved photometry during quiescence after the 2014 outburst**